\documentclass[conference]{IEEEtran}
\usepackage{cite}

\ifCLASSINFOpdf 
\usepackage[pdftex]{graphicx}
\else
\usepackage[dvips]{graphicx}
\fi

\usepackage[cmex10]{amsmath}
\usepackage{algorithmic}
\usepackage[ruled,norelsize]{algorithm2e}

\usepackage{array}

\usepackage{mdwmath}
\usepackage{mdwtab}
\usepackage{todonotes}

\usepackage{eqparbox}

\ifCLASSOPTIONcompsoc
\usepackage[caption=false,font=normalsize,labelfont=sf,textfont=sf]{subfig}
\else
\usepackage[caption=false,font=footnotesize]{subfig}
\fi

\usepackage{fixltx2e}
\usepackage{url}
\usepackage{amsfonts}
\usepackage{amssymb}
\usepackage{amsthm}
\usepackage{bm}
\usepackage[ruled]{algorithm2e}
\usepackage[utf8]{inputenc}
\usepackage{booktabs}
\usepackage{url}
\usepackage{cite}
\usepackage{epstopdf}

\newcounter{tempEquationCounter} 
\newcounter{thisEquationNumber}

\begin{document}
%


\title{5G Replicates TSN: Extending IEEE 802.1CB Capabilities to Integrated 5G/TSN Systems}


%

\author{
\IEEEauthorblockN{Adnan Aijaz}
\IEEEauthorblockA{
\text{Bristol Research and Innovation Laboratory, Toshiba Europe Ltd., Bristol, United Kingdom}\\
adnan.aijaz@toshiba-bril.com}
}   

\markboth{IEEE XYZ Magazine -- Submitted for Publication}%
{Shell \MakeLowercase{\textit{et al.}}: Bare Demo of IEEEtran.cls for Journals}
%


\maketitle
\begin{abstract}
\boldmath
The IEEE 802.1 time-sensitive networking (TSN) standards improve real-time capabilities of the standard Ethernet.  
TSN and local/private 5G systems are envisaged to co-exist in industrial environments. The IEEE 802.1CB standard provides fault tolerance to TSN systems via frame replication and elimination for reliability (FRER) capabilities. This paper presents X-FRER, a novel framework for extending FRER capabilities to the 3GPP-defined bridge model for 5G and TSN integration. The different embodiments of X-FRER realize FRER-like functionality through multi-path transmissions in a 5G system based on a single or multiple protocol data unit (PDU) sessions. X-FRER also provides enhanced replication and elimination functionality for integrated deployments. Performance evaluation shows that X-FRER empowers a vanilla 5G system with TSN-like capabilities for end-to-end reliability in integrated TSN and 5G deployments. 

\end{abstract}



%
\IEEEpeerreviewmaketitle

\section{Introduction}
\IEEEPARstart{T}{ime-sensitive networking} (TSN), which is underpinned by the IEEE 802.1 set of standards \cite{tsn}, empowers standard Ethernet to
handle stringent real-time requirements of industrial networking. TSN provides deterministic and bounded low latency and extremely low data loss which leads to guaranteed data delivery. It supports  time-critical as well as best-effort traffic over a single Ethernet network. It is expected to replace proprietary (wired) technologies in industrial systems.

On the other hand, wireless technologies have been  appearing on the industrial connectivity landscape for more than a decade. However,  the fifth-generation (5G) mobile/cellular technology has the decisive advantage of providing a unified wireless solution for diverse industrial applications. TSN will co-exist with 5G in industrial networks \cite{pvt_5G}.  

Seamless integration and converged operation of TSN and 5G systems is crucial to achieving end-to-end deterministic performance over hybrid wired/wireless domains. However, this becomes particularly challenging as the two technologies differ in capabilities and internal protocols. In the 3GPP-defined \emph{bridge model} for 5G/TSN integration \cite{3gpp_23_734}, the 5G system appears as a black box to TSN entities and handles TSN service requirements through its internal protocols.

One of the standards in the IEEE 802.1 family is the 802.1CB standard \cite{802_1CB} which provides fault tolerance to TSN systems through a frame replication and elimination for reliability (FRER) solution. It is the only TSN standard which provides protection against node and link failures. It also provides zero recovery time, i.e., protection against frame retransmissions which are detrimental to control traffic. FRER replicates each frame at source over multiple paths and provides an elimination mechanism for redundant frames at the destination.
Extending 802.1CB capabilities to integrated TSN and 5G systems becomes particularly important for end-to-end fault tolerance, especially when disjoint paths are not always available.

In literature, some studies have investigated the limitations of the IEEE 802.1CB standard.  The reliability of FRER has been investigated in \cite{Rel_FRER}. The authors show that unintentional elimination of packets can occur  when the paths chosen for a particular flow are non-disjoint. This can be mitigated through effective path selection strategies. 
The authors in \cite{Lim_FRER} identified and addressed key limitations of IEEE 802.1CB centered around recovery algorithms, sequence history, and burstiness.  Another avenue for studies on IEEE 802.1CB is simulations-based evaluation (e.g., \cite{Sim_802_1CB} and \cite{Sim_802_1CB2}). 
On the other hand, studies on end-to-end optimization of 5G systems for effectively handling TSN traffic have started to emerge (e.g., \cite{patent_virtual_TSN}); however, such studies are limited to handling of scheduled TSN traffic (IEEE 802.1Qbv). 
To the best of our knowledge, investigation of 802.1CB in integrated 5G/TSN deployments or extending FRER capabilities to the bridge model for 5G/TSN integration remain largely unexplored.

To this end, this \emph{(short)} paper presents X-FRER\footnote{Cross (\underline{X}) – technology  \underline{F}rame \underline{R}eplication and \underline{E}limination for \underline{R}eliability}, which is a novel framework for achieving FRER in integrated 5G and TSN systems. X-FRER implements a two-pronged strategy. It extends FRER-like capabilities to a 5G system through multi-path transmissions. It also provides system-level enhancements to FRER for operation in integrated deployments consisting of both 5G segments and TSN nodes and bridges. The paper also conducts performance evaluation of X-FRER by incorporating reliability theory aspects in system-level simulations.

The rest of the paper is structured as follows. Section \ref{prelim} covers some preliminaries related to TSN and 5G systems. The X-FRER framework is discussed in Section \ref{sect_fw}. Performance evaluation is presented in \ref{sect_perf}. The paper is concluded in Section \ref{sect_CR}.  


\section{Preliminaries}\label{prelim}
\subsection{Overview of TSN}
The IEEE 802.1 TSN standards can be categorized into: (a) time synchronization (e.g., 802.1AS for timing and synchronization), (b) resource management (e.g., 802.1Qcc for TSN configuration), (c) bounded low latency (e.g., 802.1Qbv for scheduled traffic), and (d) ultra-high reliability (i.e., 802.1CB for frame replication and elimination, which is the focus of this paper).  Details of these TSN standards are beyond the scope of this paper. The interested readers are referred to \cite{TSN_standards} and \cite{TSN_survey}.


\subsection{Overview of IEEE 802.1CB}
The IEEE 802.1CB FRER solution incorporates two main mechanisms: (i) replication of streams (via different paths) at the TSN talkers (source nodes), and (ii) elimination of duplicates per stream at the relay nodes or TSN listeners (destination nodes). 

\begin{figure}
\centering
\includegraphics[scale=0.47]{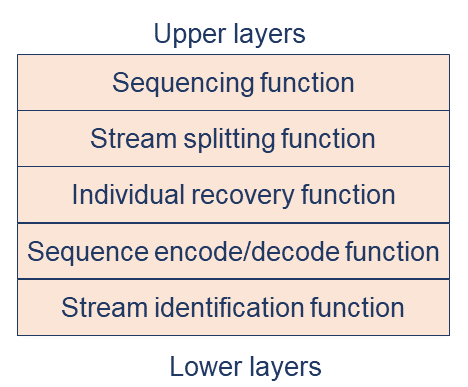}
\caption{Key functionalities of the FRER protocol stack.}
\label{stack_FRER}
\end{figure}

FRER provides five key functions as shown in Fig. \ref{stack_FRER}. Note that not all the functions are required in all protocol stacks. The sequence generation function generates a sequence number for each frame of a stream. The stream splitting function generates \(K\) copies for each frame of a stream to be forwarded via \(K\) disjoint/distinct paths.  The sequence encoding function assigns the generated sequence number to each frame using the redundancy tag (R-TAG) field of the Ethernet frame. The sequence decoding function extracts the sequence number from a received frame. The sequence recovery stage either accepts or discards a received frame. The sequence recovery function operates at the level of compound streams, i.e., frames received from all the paths. The individual recovery function operates at the level of a member stream, i.e., frames received from a single path. The stream identification function identifies the stream to which the frame belongs to. This is achieved through source or destination MAC addresses or VLAN ID in the Ethernet frame. 

\subsection{5G Quality-of-Service Framework}
Quality-of-Service (QoS) in a 5G system is enforced at the \emph{QoS flow} level. Each QoS flow is identified by a unique QoS flow ID (QFI). The data between the UE and the core network is transferred in a protocol data unit (PDU) session. Multiple QoS flows can be carried in a single PDU session. The 5G system supports both IP and Ethernet type PDU sessions. 

\section{X-FRER Framework} \label{sect_fw}

\subsection{Network Model}
\begin{figure}
\centering
\includegraphics[width=1\columnwidth]{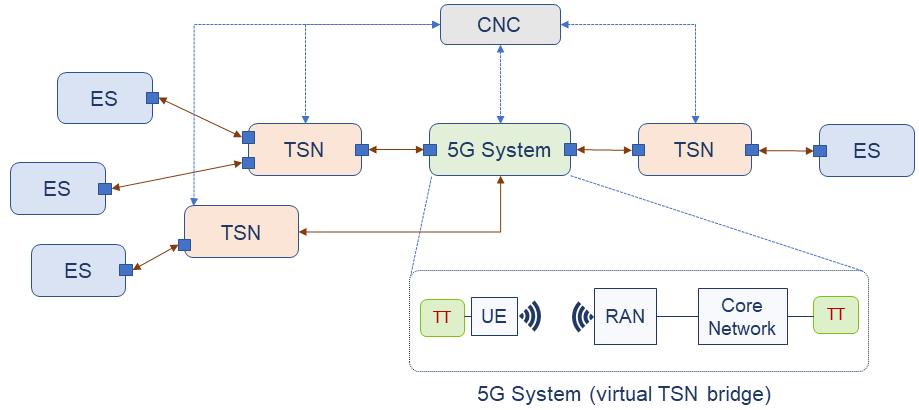}
\caption{Illustration of the network model for integrated 5G and TSN systems.}
\label{net_model}
\end{figure}

We consider a network model depicting an integrated 5G and TSN system which is illustrated in Fig. \ref{net_model}. It consists of TSN end stations (ESs) which can be talkers or listeners, TSN bridges, a 5G system, and a centralized network configuration (CNC) entity for management of the integrated system. The 5G system consists of at least one user equipment (UE), a radio access network (RAN) comprising one or more base stations (gNBs) and a 5G core network. We assume that the TSN system implements FRER functionality.

The network model is aligned with the 3GPP-defined \emph{bridge model} for 5G and TSN integration.  X-FRER extends FRER capabilities to a 5G network through multi-path transmissions which provide FRER-like functionality. Fig. \ref{mp_5G} shows the concept of multi-path transmissions in a 5G network. Such multi-path transmissions can be realized via multiple PDU sessions or within a single PDU session.

\subsection{FRER-like Functionality via Multiple PDU Sessions}
\begin{figure}
\centering
\includegraphics[width=1\columnwidth]{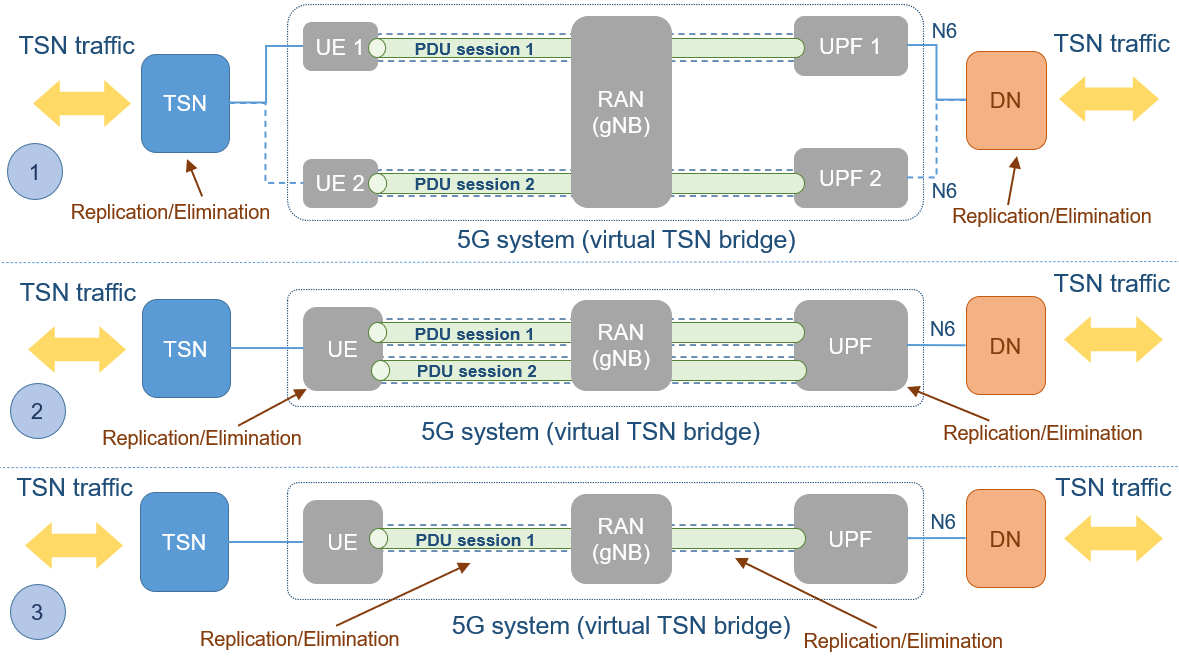}
\caption{Illustration of multi-path transmissions in a 5G system.}
\label{mp_5G}
\end{figure}

\begin{figure}
\centering
\includegraphics[width=1\columnwidth]{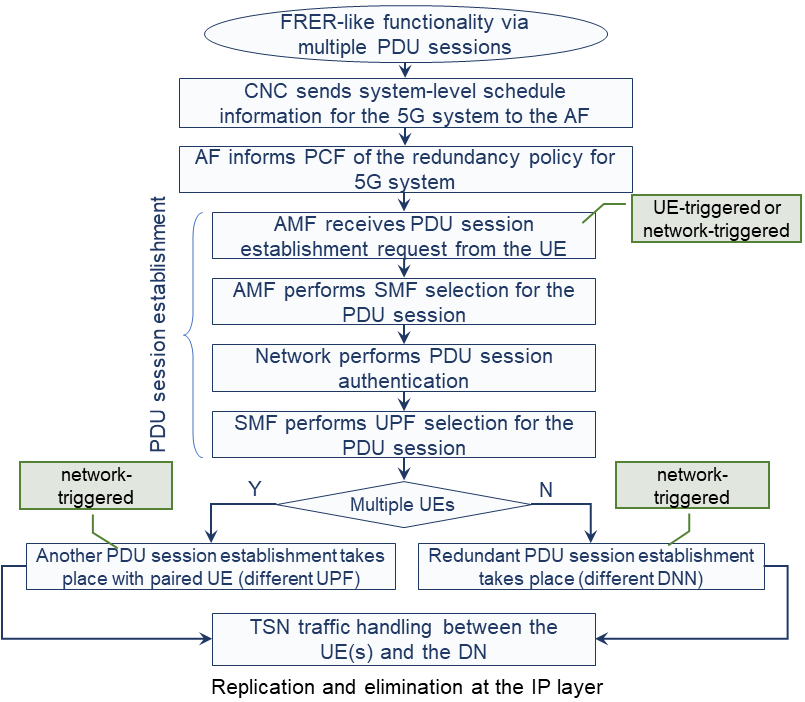}
\caption{FRER-like functionality with multiple PDU sessions.}
\label{FRER_multi}
\end{figure}

In some embodiments FRER-like functionality is realized through multiple PDU sessions and multiple UEs. The scenario and protocol operation are shown in Fig. \ref{mp_5G} and Fig. \ref{FRER_multi}, respectively. Initially, the CNC sends system-level schedule information for the 5G system to the application function (AF). The AF informs the policy control function (PCF) of the redundancy policy for the 5G system. The system-level schedule informs the 5G system of the time window during which the TSN traffic is expected to arrive at the 5G system. The redundancy policy indicates what type of TSN traffic handling mechanism is supported by the 5G system, from the perspective of frame replication and elimination.
In this scenario, multiple UEs handling TSN traffic are grouped together. The network recognizes these UEs through their IMSI or any other unique identifier. We illustrate protocol operation with two UEs. The access and mobility management function (AMF) receives a PDU session establishment request from one of the UEs. This PDU session establishment can be triggered by the UE or the network. The AMF selects a session management function (SMF) for the PDU session. The core network performs necessary functions related to PDU session authentication/authorization. After this, the SMF selects a user-plane function (UPF) for the PDU session. Once the PDU session establishment is complete for one UE, the network triggers PDU session establishment for the second (paired) UE. The previous steps are repeated with the exception that a different UPF is selected for the PDU session. Therefore, two different paths are established within a 5G system to handle TSN traffic.
As shown in Figure 8, the TSN traffic can enter the 5G system from the UE side or the network side. In either case, it is transmitted via two different paths, each having its own PDU session with separate UEs and UPFs as end points. The TSN traffic entering the 5G system is replicated at the IP layer (or Layer 2 in case of Ethernet type PDUs). Frame/Packet replication and elimination takes place at the data network (DN) or the TSN node. 

In some embodiments, multiple PDU sessions are established with a single UE. This scenario is shown in Fig. \ref{mp_5G}. Similar steps are followed in this case as shown in Fig. \ref{FRER_multi}; however, once a PDU session is successfully established, a redundant PDU session establishment procedure with the same UE is triggered by the network. Two different PDU sessions with the same UE are established via two different data network names (DNNs). In this case, Frame/Packet replication and elimination takes place at the IP layer, at the UPF or the UE.

\subsection{FRER-like Functionality in a Single PDU Session}
\begin{figure}
\centering
\includegraphics[width=1\columnwidth]{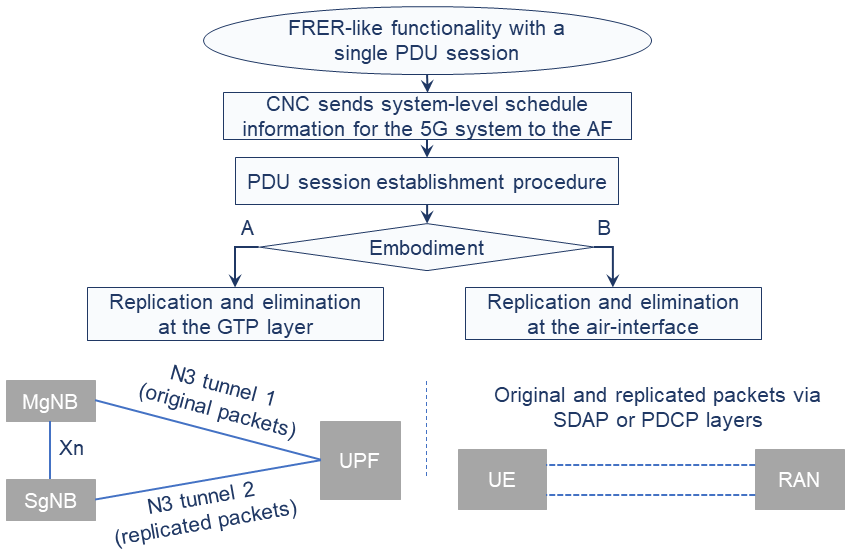}
\caption{FRER-like functionality with a single PDU session.}
\label{FRER_single}
\end{figure}

In some embodiments, FRER-like functionality in a 5G system is achieved via a single PDU session. This scenario and the protocol operation is shown in Fig. \ref{mp_5G} and Fig. \ref{FRER_single}, respectively. Once a PDU session establishment procedure is complete, frame/packet replication/elimination is handled via two different embodiments.

In the first embodiment (Embodiment A), frame/packet replication and elimination takes place between the UPF and the RAN at the GPRS tunnelling protocol (GTP) layer. The UPF establishes two different N3 tunnels for TSN traffic handling via two different paths. On the RAN side, the two N3 tunnels can have separate end points enabled by dual-connectivity techniques where two different gNBs are used: a master gNB (MgNB) and a secondary gNB (SgNB). The two gNBs are connected via the Xn interface.

\begin{figure}
\centering
\includegraphics[width=1\columnwidth]{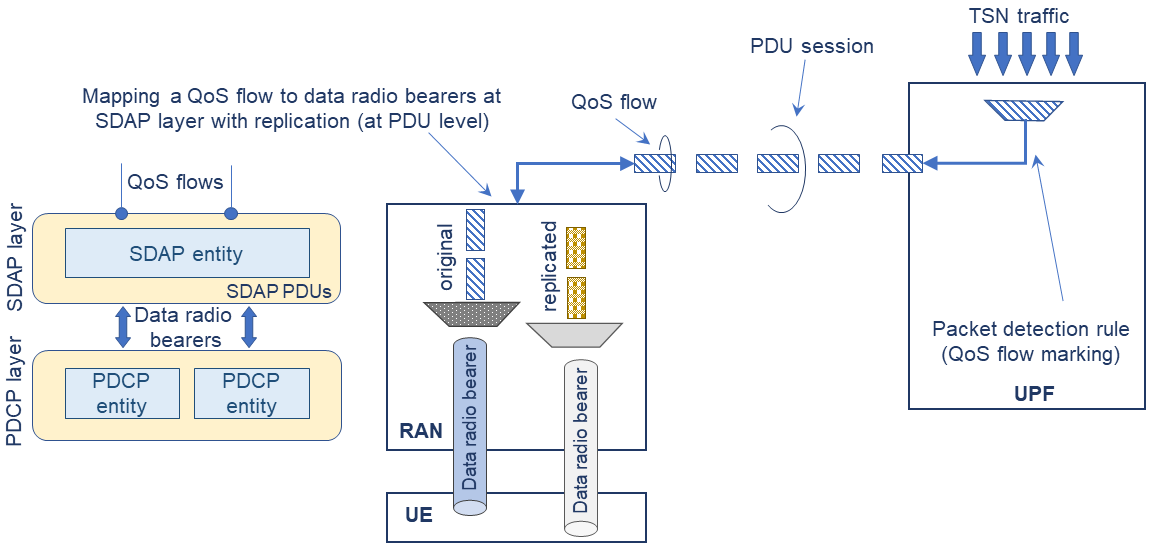}
\caption{Illustration of replication at the SDAP layer.}
\label{FRER_SD}
\end{figure}

In the second embodiment (Embodiment B), frame/packet replication and elimination takes place between the UE and the RAN at different layers of the air-interface.
In some scenarios, replication and elimination takes place at the service data adaptation protocol (SDAP) layer which handles mapping of QoS flows to data radio bearers. As depicted in Fig. \ref{FRER_SD}, the SDAP layer duplicates the SDAP PDUs and adds another PDCP entity such that the original and replicated PDUs are sent via two different data radio bearers (DRBs).

\begin{figure}
\centering
\includegraphics[width=1\columnwidth]{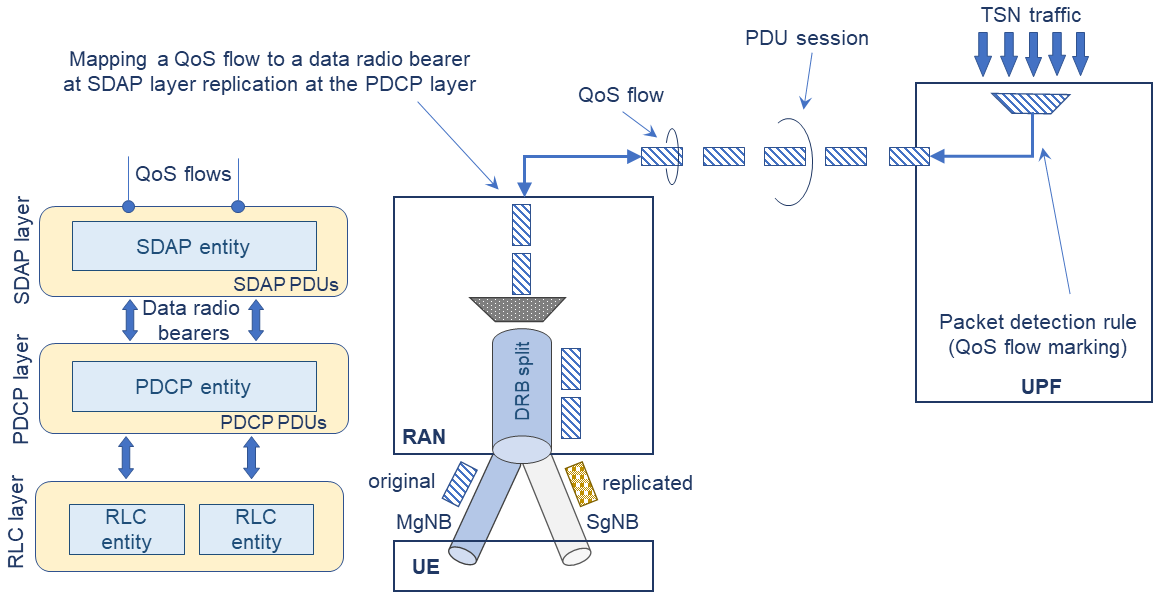}
\caption{Illustration of replication at the PDCP layer.}
\label{FRER_PD}
\end{figure}

In some scenarios, frame replication and elimination takes place at the packet data convergence protocol (PDCP) layer. As illustrated in Fig. \ref{FRER_PD}, the PDCP layer replicates the PDCP PDUs. It adds another radio link control (RLC) entity and performs a bearer split operation such that the original and the replicated PDUs are sent via two different paths: one via the MgNB and the other via the SgNB.

In some scenarios, both Embodiment A and Embodiment B can be used simultaneously.

\subsection{FRER-like Functionality with Multiple Techniques}
The FRER-like functionality can also be achieved by combining multiple techniques. Multiple PDU sessions can be used while replication/elimination techniques can be utilized within a single PDU session.

\begin{figure}
\centering
\includegraphics[width=1\columnwidth]{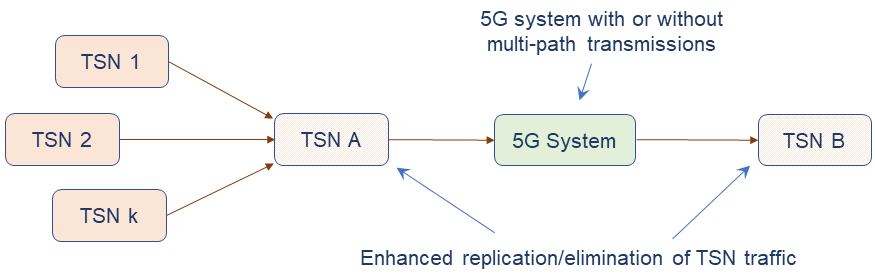}
\caption{System model for illustrating system-level enhancements provided by X-FRER.}
\label{FRER_enh}
\end{figure}

\subsection{System-level Enhancements}
X-FRER provides system-level enhancements for FRER operation in integrated 5G and TSN deployments. Such enhancements are necessary as disjoint end-to-end paths might not be available when a 5G system is shared among multiple TSN systems. Consider the system model shown in Fig. \ref{FRER_enh}. X-FRER provides enhanced replication/elimination functionalities for TSN nodes transmitting/receiving traffic to/from a 5G system.

In some embodiments, an FRER-compliant TSN node transmitting traffic toward a 5G system dynamically eliminates received replicated packets from multiple paths based on the level of redundant paths available in the 5G system. A TSN node transmitting traffic toward a 5G system \emph{allows} received replicated frames from \(k\) paths with probability \(1/\Theta\) where \(\Theta\) is the average number of paths available in a 5G system for TSN traffic and it is calculated as
\begin{equation}
\label{eq_paths}
[\mathcal{P}_{\textrm{UE-RAN}} + \mathcal{P}_{\textrm{RAN-Core}} ]/2
\end{equation}
where \(\mathcal{P}_{\textrm{UE-RAN}}\) denotes the total number of available paths for redundant traffic between the UE and the RAN, and \(\mathcal{P}_{\textrm{RAN-Core}}\) is the total number of available paths for redundant traffic between the RAN and the core network. Note that, in case of 2 independent PDU sessions (either with a single UE or multiple UEs), the average number of paths is 2. In case of a single PDU session with replication/elimination at the air interface or the N3 interface, the average number of paths is 1.5. In case of no multi-path transmissions supported by the 5G system, the average number of paths is 1.

In some embodiments, a TSN node receiving traffic from a 5G system \emph{replicates} received frames (over \(k\) available paths) with probability \(1/\Theta\) where \(\Theta\) is given by \eqref{eq_paths}.

The information about the average number of paths in a 5G system can be shared with the TSN nodes via the CNC.

\subsection{Key Advantages}
X-FRER primarily provides FRER-like functionality for a 5G system and optimizes multi-hop operation in integrated deployments. Some of the additional benefits of X-FRER are as follows.
\begin{itemize}
\item It provides resilience against wireless imperfections and different types of failures within a 5G system.

\item It provides deterministic capabilities for a 5G system.

\item It improves resilience of FRER in scenarios where disjoint paths are not available.

\item It provides more efficient utilization of resources in integrated 5G and TSN environments. 

\end{itemize}

\section{Performance Evaluation} \label{sect_perf}
\begin{figure}
\centering
\includegraphics[width=1\columnwidth]{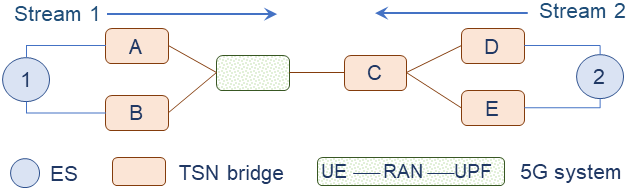}
\caption{Topology for the performance evaluation of X-FRER.}
\label{top}
\end{figure}

We conduct performance evaluation of X-FRER through system-level simulations incorporating reliability theory aspects which consider node and link failures. We consider the topology shown in Fig. \ref{top} which comprises an integrated system of TSN ESs, TSN bridges, and 5G system with UE, RAN/gNB, and a UPF. Note that this topology lacks end-to-end disjoint paths. We consider two streams\footnote{For stream 1, the hops are labelled as follows: hop 1 (ES1 to TSN bridges A and B), hop 2 (TSN bridges A and B to the UE in 5G system), hop 3 (UE to RAN/gNB), hop 4 (RAN/gNB to UPF), hop 5 (UPF to TSN bridge C), hop 6 (TSN bridge C to TSN bridges D and E), hop 7 (TSN bridges D and E to ES 2). Stream 2 follows the reverse path of stream 1, and therefore, the hops are labelled as follows: hop 1 (ES2 to TSN bridges D and E), hop 2 (TSN bridges D and E to TSN bridge C), hop 3 (TSN bridge C to UPF in 5G system), hop 4 (UPF to RAN/gNB), hop 5 (RAN/gNB to UE), hop 6 (UE to TSN bridges A and B), and hop 7 (TSN bridges A and B to ES 1). } for performance evaluation: \textbf{stream 1 (ES1 to ES2)} and \textbf{stream 2 (ES2 to ES1)}. 
The simulation results are based on exchange of 1 million frames between ESs in each stream.

The node failure probabilities used in our evaluation include \(\mathcal{N}_B=10^{-5}\), \(\mathcal{N}_{UE}=\mathcal{N}_{gNB}=\mathcal{N}_{UPF}=10^{-4}\), where 
\(\mathcal{N}_B\), \(\mathcal{N}_{UE}\), \(\mathcal{N}_{gNB}\), \(\mathcal{N}_{UPF}\) denote the node failure probability for the TSN bridges, the UE, the gNB/RAN, and the UPF, respectively. The link failure probabilities include \(\mathcal{L}_{TSN}=10^{-4}\), \(\mathcal{L}_{u}=10^{-2}\), \(\mathcal{L}_{N3}=10^{-3}\), such that \(\mathcal{L}_{TSN}\), \(\mathcal{L}_{u}\), \(\mathcal{L}_{N3}\) denote the link failure probabilities for the TSN links, the air-interface between the UE and the gNB/RAN, and the N3 interface between the gNB and the UPF, respectively. We assign higher reliability to TSN bridges and links as compared to the 5G system. The N3 interface is assumed to be Ethernet without TSN capabilities; hence its lower link reliability as compared to the TSN links.

\begin{figure}
\centering
\includegraphics[width=1\columnwidth]{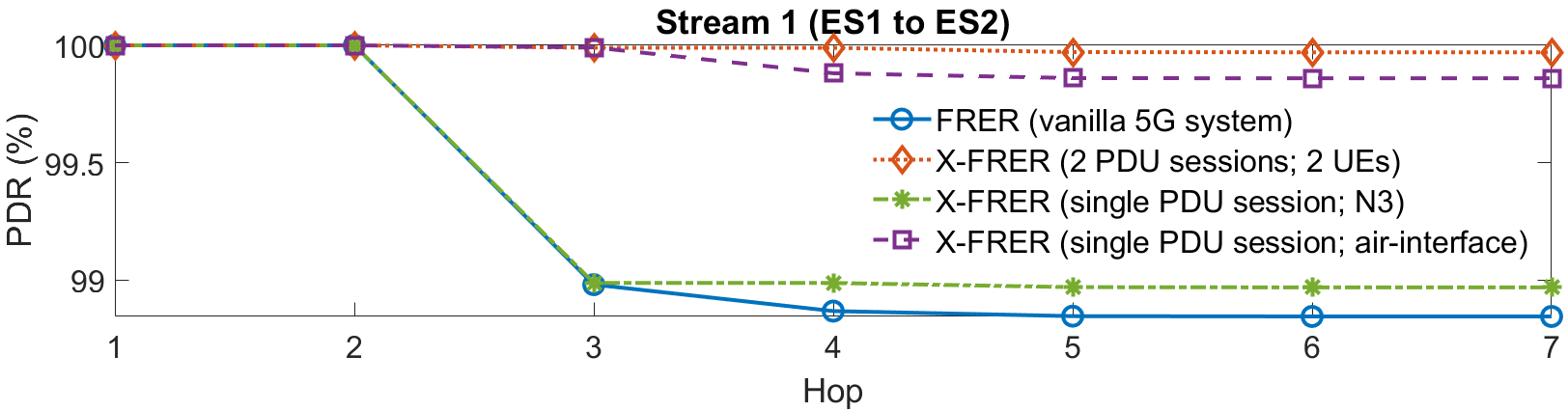}
\caption{Reliability performance of X-FRER for stream 1 in different scenarios.}
\label{perf1}
\end{figure}

Fig. \ref{perf1} captures the end-to-end reliability performance of the integrated system in the form of packet/frame delivery ratio (PDR) in different operational scenarios for stream 1. The reliability performance degrades to \(98.98\%\) at the third hop when the TSN traffic passes through the air-interface of the 5G system. In case of vanilla 5G system, where FRER is only used by TSN bridges, the end-to-end reliability performance experiences degradation to \(98.84\%\). However, X-FRER provides crucial performance improvement. The highest end-to-end reliability performance of \(99.97\%\) is achieved with two independent paths in a 5G system, enabled by two separate PDU sessions. X-FRER with a single PDU session also provides significant performance improvement, particularly with multi-path transmissions on the air-interface (i.e., to \(99.86\%\)).

\begin{figure}
\centering
\includegraphics[width=1\columnwidth]{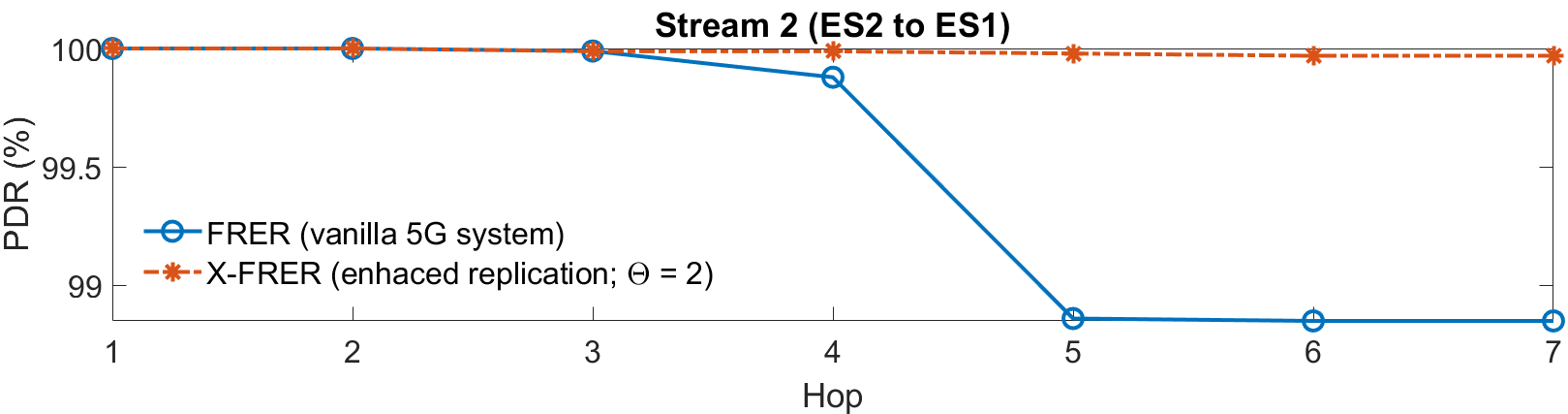}
\caption{Reliability performance of X-FRER for stream 2 with enhanced replication functionality.}
\label{perf2}
\end{figure}

Fig. \ref{perf2} captures the performance benefits of system-level enhancements in X-FRER for end-to-end reliability in case of stream 2. With vanilla 5G system (without multi-path transmissions) and no system-level enhancements for X-FRER, the end-to-end reliability performance experiences degradation at the fourth and fifth hops, i.e., \(99.87\%\) and \(98.85\%\), as the TSN traffic passes through the 5G system (from the UPF to the UE). However, with enhanced replication functionality at TSN bridge C, which is forwarding TSN traffic to the 5G system, the end-to-end reliability is significantly improved from \(98.84\%\) to \(99.96\%\).

\section{Concluding Remarks}\label{sect_CR}
Integration of 5G and TSN systems is an important step in the evolution of industrial networks. Achieving TSN-like functionality in a 5G system, without native support for TSN protocols, is a promising approach for converged operation of the two systems. This paper proposed X-FRER which extends IEEE 802.1CB FRER capabilities to a 5G system along with system-level enhancements for integrated deployments. FRER-like functionality in a 5G system is achieved via frame replication/elimination with multiple PDU sessions or within a single PDU session. 

Performance evaluation in a reference integrated multi-hop 5G/TSN system with non-disjoint paths, while considering node and link failures, demonstrates the effectiveness of X-FRER in terms of guaranteeing end-to-end reliability. X-FRER empowers a vanilla 5G system to act like a virtual TSN bridge in integrated 5G/TSN deployments.

\bibliographystyle{IEEEtran}
\bibliography{bibliography.bib}

\end{document}